\documentstyle[12pt]{article}
\begin{document}
\begin{titlepage}
\begin{flushright}
IMSc-92/20\\
\today\\
\end{flushright}
\begin{center}
\vspace{0.4cm}
\Large {\bf Can Polarised Drell-Yan Shed More Light On The
Proton Spin ?}
\vspace{1.2cm}
\normalsize

\centerline{\large\bf Prakash Mathews$^{\dag}$ and V. Ravindran$^{\ddag}$}
\centerline{\em The Institute of Mathematical Sciences,}
\centerline{\em C.I.T. Campus,  Madras, 600 113, INDIA.}
\vspace{.3cm}

\centerline{\bf Abstract}
\begin{quote}
We analyse polarised Drell-Yan process using the factorisation method
and derive operator definitions for polarised parton distribution
functions.  We demonstrate that a factorisation analogous to that in the
unpolarised Drell-Yan case holds in this process.  We study the leading
order gluonic contribution to the
first moment of polarised Drell-Yan function and show that it is consistent
with results obtained from polarised deep inelastic scattering.

\end{quote}
\end{center}
\vspace{.5cm}
\begin{flushleft}
e-mail address:\\
\nopagebreak
$\dag$ prakash@imsc.ernet.in\\
\nopagebreak
$\ddag$  ravi@imsc.ernet.in\\
\end{flushleft}
\end{titlepage}

The post Electron Muon Collaboration (EMC) \cite{EMC} era has seen a
revival of interest in the theoretical study of the polarised structure
functions of the proton.  The EMC experiment (polarised deep inelastic
$\mu P$  scattering) gave some surprising results regarding the quark
spin contribution to the proton spin, namely, that it is negligibly small.
The most
appealing interpretation of this result is that not only the quarks but
the gluons also can
contribute to the proton spin, to the lowest order. This is due to the
fact that  the product of
strong coupling constant, $\alpha_s$,
and the first moment of polarised gluon distribution function,
$\Delta g(x)$, is of order $\alpha_s^0$ \cite{LL,AR}.
In the parton model, using the electromagnetic probe to evaluate the
 gluonic coefficient functions to
leading order, one needs a prescription to regulate mass singularities
present, and depending on the prescription the gluons may or may not
contribute to the first moment of the polarised proton structure
function, $g^{\mu P}_1(x)$ \cite{AR,ET}.  Even
if one uses the weak probe to evaluate the gluonic coefficient function,
one would draw similar conclusions for the moments of the weak structure
functions $g_{1,3,4}(x)$ \cite{PR}.  On the other hand, in the
operator product expansion  \cite{K} and factorisation method
\cite{CSS,BQ}, the gluonic
coefficient function is free of mass singularities and its first
moment vanishes identically giving no gluonic contribution to the first
moment of $g^{\mu P}_1(x)$ to order $\alpha_s$ \cite{BQ}.  Hence, using
the factorisation
method, in a class of deep inelastic scattering (DIS) processes probed by
electromagnetic \cite{BQ} and weak currents \cite{PR1}, one finds that
the gluonic contribution to the first moments of polarised structure
functions, $g_1^{\mu P}(x)$ and $g_{1,3,4}(x)$, is zero.  Under these
circumstances, it is
therefore instructive to analyse the
polarised Drell-Yan (DY) processes as possible sources to measure the
gluonic contribution to proton spin.  Phenomenologically \cite{IB},
this process
has drawn much attention after the EMC experiment.  Further, the
experimental possibilities
\cite{RMC} are also bright with the recent technical developments in the
acceleration of polarised protons, using the so-called Siberian snakes,
which facilitate the polarised proton-proton collision at collider
energies.

In this letter we analyse the polarised DY process using the
factorisation method and show that the polarised DY
distribution function $\widetilde {W}(\tau,q^2)$  can be
expressed in terms of known operators available in the theory.
Further including the appropriate gluonic operators we write a
factorisation theorem for
polarised DY process and show that the mass singularities
factorise leaving the gluonic coefficient functions free of mass
singularities to order $\alpha_s$.

Consider the longitudinally polarised DY process $P_1(p_1)^{\uparrow}+
P_2(p_2)^{\uparrow \downarrow} \rightarrow
\gamma^*(q)+X(P_X)$, where $\gamma^*$ is the virtual hard photon.  The
cross section of the process is
\begin{eqnarray}
\tilde \sigma & = & {1 \over {2K(2 \pi)^3}}(-g_{\mu \nu}) \int d^4 q~
\delta (q^2-Q^2) ~\theta (q^0)~ \widetilde W^{\mu \nu},
\label{eq:E1}
\end{eqnarray}
where $K$ is the flux factor, $Q^2$ is mass of the virtual hard photon,
the sum over photon polarisations
yields the factor $-g^{\mu \nu}$ and $\widetilde W^{\mu \nu}$ is the
Fourier transform of
product of hadronic currents sandwiched between two polarised proton states.
The polarised DY function $\widetilde W_{P_1 P_2}(\tau,q^2)$ is defined as
\begin{eqnarray}
\widetilde W_{P_1 P_2}(\tau ,q^2)& = & - {1 \over {(2 \pi)^3 }} \int d^4 q~
\delta
(q^2-Q^2)~ \theta (q^0) \int d^4 x ~e^{-iq.x}
\nonumber\\
 & & <p_1 s_1;p_2 s_2 \vert~ J_\mu (x)~~ J^\mu (0)~ \vert p_1 s_1;p_2 s_2
>_{sp}.
\label{eq:E2}
\end{eqnarray}
We choose the centre of momentum light cone frame where $p_1=(p_1^+,0,0_
\perp)$, $p_2=(0,p_2^-,0_\perp)$, $q=(q^+,q^-,0_\perp)$, $x=(x^+,x^-,x_\perp)$
and the light cone variables are defined as $p^{\pm} =(p^0 \pm p^3)/{\sqrt 2}$.
Using the DY limit $p_1^+,p_2^- \rightarrow \infty$, $q^+,q^- \rightarrow
\infty
$ with $\tau ={q^+q^-/p_1^+p_2^-}$ ($\tau = Q^2/s$, where {\it s} is the cm
energy) fixed, we can express the product of hadron
currents in terms of the parton currents which are made up of parton
field operators coming from two different protons.
In order to proceed, we need to express the current-current product so
that the dependences on the two proton momenta separate out.  We do this
by a Fierz transformation whereby the product of local bilinear (with
respect to two different proton states) operators is expressed as
sum of bilocal linear operators.
Hence the DY function can be neatly separated out in terms of two
proton states and we obtain,
\begin{eqnarray}
\widetilde W_{P_1 P_2}(\tau,q^2) & = & \sum _i {e_i^2 \over {24 \pi}} \int
{dq^+
\over q^+} {dq^- \over q^-} \delta (1 - {Q^2 \over {2 q^+ q^-}})
\int {dx^+ dx^-} e^{-iq^+x^-} e^{-iq^-x^+}
\nonumber\\
& & \left [ <p_1 s_1 \vert~ \overline \psi _a (x^+,x^-)~\gamma _\mu~\gamma _5
\psi _a (0) ~\vert p_1 s_1>~\times \right.
\nonumber\\
& & \left. <p_2 s_2 \vert ~\overline \psi _b (0)~
\gamma ^\mu ~\gamma _5 \psi _b (x^+,x^-)~ \vert p_2 s_2> +
(a  \leftrightarrow b) \right ]~~,
\label{eq:E3}
\end{eqnarray}
where {\it i} runs over different quark flavours, and
the scalar, vector and pseudoscalar terms
which appear in the Fierz identity do not contribute for the polarised DY
process.  From the asymptotic behaviour of Fourier transform of singular
functions, it is easy to see that the dominant contribution comes
from the region near the singularities \cite{LW}.  Hence, using
$q^+=x_a p_1^+$ and $q^-=x_b p_2^-$, we have
\begin{eqnarray}
\widetilde W_{P_1 P_2}(\tau,q^2) & = & \sum _i  \int {dx_a \over x_a}
{dx_b \over x_b}~{{2 \pi e_i^2}\over 3} ~\delta (1 - {\tau \over {x_a x_b}})
\nonumber\\
& & \left [ {1 \over {4 \pi}} \int {dx^-} e^{-ix_a p^+_{1} x^-}
<p_1 s_1 \vert ~ \overline \psi _a (x^-) ~ \gamma ^+ ~ \gamma _5
\psi _a (0) ~ \vert p_1 s_1> ~\times \right.
\nonumber\\
& & \left. {1 \over {4 \pi}} \int d x^+ e^{-i x_b p^-_{2} x^+}
<p_2 s_2 \vert ~ \overline \psi _b (0) ~ \gamma ^- ~ \gamma _5 ~ \psi _b (x^+)
\vert p_2 s_2> + (a \leftrightarrow b) \right ],
\label{eq:E4}
\end{eqnarray}
where we have also used the fact that $ <p^{\pm}s\vert ~ \overline \psi
{}~\gamma^{\mp}~ \gamma_5~ \psi ~\vert p^{\pm} s>   =  0$.
The above expression provides neat separation of short and long distance
physics.  By noting that the Dirac delta function corresponds to the
$q,\bar q \rightarrow \gamma^*$ subprocess, the terms in the bracket can
be identified with the polarised quark and antiquark distribution
functions of the proton, which are consistent with the DIS polarised
distribution functions \cite{BQ} and hence confirms the universality of
the parton distribution functions.

To make the definitions of the parton distribution functions
gauge invariant, one has to insert the path
ordered exponential given below.  With this, one can generalise the
above expression
by replacing delta function by hard scattering coefficient function
$\widetilde H_{ab}$ and adding appropriate gluonic operators of the
theory.  We thus have

\begin{eqnarray}
\widetilde W_{P_1 P_2}(\tau,q^2) & = & \sum_{a,b} \int_{\tau}^{1}{dx_a \over
x_a}
\int_{\tau \over x_a}^{1} {dx_b \over x_b}~ f_{\Delta a \over P_1} (x_a,\mu^2)
{}~\widetilde H_{ab} ({\tau \over {x_a x_b}},{ q^2 \over \mu^2})~
f_{\Delta b \over P_2} (x_b,\mu^2) +
\nonumber\\
& & (a \leftrightarrow b) + \cdots ,
\label{eq:E5}
\end{eqnarray}
where $\mu^2$ is the factorisation scale, $\widetilde W_{P_1 P_2} = {1 \over 2}
(W_{++} - W_{+-}),
\,\ f_{\Delta a/ A}$ is the polarised probability distribution of the
parton $a (\equiv q_i,{\bar q_i},g)$ in the polarised hadron target A,
defined as $f_{\Delta a/A_+ } = (f_{a_+/A_+} - f_{a_-/A_+})$ and $\widetilde H
_{ab} $ is the polarised hard scattering coefficient defined as
$\widetilde H_{ab}= {1\over 2} (H_{a_+ b_+} - H_{a_+ b_-})$
where the $+(-)$ sign denotes the polarisation of the proton or parton along
(opposite to) the beam axis.  The ellipses in eq.(\ref{eq:E5}) represent the
higher twist contributions.  The gauge invariant definitions of polarised quark
and antiquark distributions in the proton are identified using eq.(\ref{eq:E4}
) in
terms of the connected matrix elements of the bilocal operators as

\begin{eqnarray}
f_{\Delta q \over A}(x) & = & {1\over 4\pi} \int d\xi^-
{e^{-ix\xi^-p^+}}\left[ <p,s\vert~\overline { \Psi}_a(0,\xi^-,0_\perp)~
\gamma^+
{}~\gamma_5~{\cal G}^a_b~ \Psi^b(0)~ \vert p,s>_c \right],
\label{eq:E6}\\
f_{{\Delta \bar q} \over A}(x) & = & {1\over 4\pi} \int d\xi^-
{e^{-ix\xi^-p^+}} \left[
 <p,s\vert~\overline {\Psi}_a(0) ~\gamma^+~\gamma_5
{\cal G}^{a \dagger}_b~  \Psi^b(0,\xi^-,0_\perp)~ \vert p,s>_c \right].
\label{eq:E7}
\end{eqnarray}
In terms of the appropriate gluonic operators in the theory the gluonic
distribution function is defined as
\begin{eqnarray}
f_{\Delta g \over A}(x) & = & {i\over 4\pi x p^+} \int d\xi^- {e^{-ix\xi^-p^+}}
\left[ <p,s\vert ~F_a^{+ \mu}(0,\xi^-,0_\perp)~ {\cal G}^a_b~
{}~\widetilde F_{\mu}^{+ b}(0) ~\vert p,s>_c - \right.
\nonumber\\
& & \left.  <p,s\vert ~F_a^{+ \mu}(0) ~{\cal G}^a_b ~\widetilde
 F_{\mu}^{+ b}(0,\xi^-,0_\perp)~ \vert p,s>_c \right],
\label{eq:E8}
\end{eqnarray}
where $F^{\mu \nu}_a$ is the gluon field strength operator, ${\cal G}^a{}_b =
{\cal P} exp~[i g \int _{0}^{\xi^-} d\zeta^- A^+(0,\zeta^-,0 _\perp)]~^a_b$,
$~\cal P$ denoting the path ordering of the gluon field operators $A^\mu$.
For known targets like quarks and gluons the
distribution functions are normalised as
\begin{eqnarray}
f_{{\Delta q + \Delta \bar q} \over a(h)}(z) &=& h~ \delta (1-z)
{}~\delta_{a,(q,\bar q)},
\label{eq:E9} \\
f_{\Delta g \over a(h)}(z) & = & h ~\delta (1-z) ~\delta_{a,g}~,
\label{eq:E10}
\end{eqnarray}
where $h  = \pm 1$ denotes the helicity of the incoming parton.  In general,
the above
matrix elements cannot be calculated within the domain of perturbative QCD,
owing to the complex nature of the proton target.  But these are calculable for
specific parton targets like quarks and gluons.  Leading order corrections to
these distributions are both ultraviolet (UV) and mass sensitive.
These can be regularised and renormalised in any scheme.  We use dimensional
regularisation to regularise the UV divergences and $\overline {MS}$ scheme to
renormalise these distributions.  This choice is preferred as it ensures gauge
invariance and relativistic invariance although it destroys scale invariance.
The mass singularities are avoided by considering on-shell quarks in
the cut quark loop and off-shell gluons.  Unlike the distribution functions
$f_{\Delta a/A}$, the hard scattering coefficients are determined in the theory
without any reference to the target.  These coefficients may also be mass and
UV
sensitive and should be regularised in the same scheme.

The $n^{th}$ moment of the DY function can be
related to the $n^{th}$ moment of the distribution functions and the hard
scattering coefficients as
\begin{eqnarray}
\widetilde W_{A B}(n,q^2) = \sum_{ab} f_{\Delta a\over A}(n,\mu^2)
{}~\widetilde H_{ab}
\left( n,{q^2\over \mu^2}\right )~ f_{\Delta b \over B}(n, \mu^2) ,
\label{eq:E11}
\end{eqnarray}
where $f(n) = \int_{0}^{1} x^{n-1} f(x) dx$.
Here we can easily identify $f_{\Delta a/A}(n,\mu^2)$
with the matrix elements of the twist two local operators.

One of the interesting offshoots of the EMC interpretation is that
$\alpha_s \Delta g$  is of order $\alpha_s^0$.  Hence the leading order
contribution to the DY process comes from the following subprocesses $q
\bar q \rightarrow \gamma^*$, $g(q,\bar q) \rightarrow \gamma^* (q,\bar
q)$ and $gg \rightarrow {\gamma^ *} q {\bar q}$.
We therefore use the factorisation theorem to calculate the gluonic
contribution to the spin dependent DY function to
order $\alpha_s$ and the quark contribution to lowest order.  To this
order, the quantity of interest is
\begin{eqnarray}
\widetilde W_{P_1 P_2}(\tau,q^2)& = &{\sum_{i}} \int
_{\tau}^{1} {dx_a\over x_a} \int_{\tau \over x_a}^1 {dx_b \over x_b}
f_{\Delta q_i /P_1}(x_a,\mu^2)~
\widetilde H_{q_i \bar q_i}( {\tau \over {x_a x_b}} ,{q^2 \over \mu^2})
{}~f_{\Delta \bar q_i /P_2}(x_b,\mu^2) +
\nonumber\\
& &{\sum_{i}} \int
_{\tau}^{1} {dx_a\over x_a} \int_{\tau \over x_a}^1 {dx_b \over x_b}
{}~f_{\Delta q_i /P_1}(x_a,\mu^2)
{}~\widetilde H_{q_i g}( {\tau \over {x_a x_b}},{q^2 \over \mu^2} )
{}~f_{\Delta g /P_2}(x_b,\mu^2) +
\nonumber\\
& &{\sum_{i}} \int
_{\tau}^{1} {dx_a\over x_a} \int_{\tau \over x_a}^1 {dx_b \over x_b}
{}~f_{\Delta \bar q_i /P_1}(x_a,\mu^2)
{}~\widetilde H_{\bar q_i g}({\tau \over {x_a x_b}} ,{q^2 \over \mu^2})
{}~f_{\Delta g /P_2}(x_b,\mu^2) +
\nonumber\\
& & (a \leftrightarrow b).
\label{eq:E12}
\end{eqnarray}
Using the Born diagram (fig.1), we get
\begin{eqnarray}
\widetilde {H}_{q \bar q}(\hat \tau) & = & {{2 \pi e_i^2} \over 3}~ \delta
(1-\hat \tau).
\label{eq:E14}
\end{eqnarray}
To evaluate the gluonic contribution to the DY function
$\widetilde W_{P_1 P_2}(\tau,q^2)$ we replace the proton targets by quark and
gluons and use eq.(\ref{eq:E6}), eq.(\ref{eq:E7}), eq.(\ref{eq:E8}) and
eq.(\ref{eq:E12}) and express the resulting `compton' subprocess to
order $\alpha_s$ as
\begin{eqnarray}
\widetilde W_{qg}(\hat \tau,q^2)& = &{\sum_{i}} \int
_{\hat \tau}^{1} {dy_a\over y_a} \int_{\hat {\tau} \over y_a}^1 {dy_b \over
y_b}
{}~f_{\Delta \bar q_i /g}(y_a,\mu^2)
{}~\widetilde H_{q_i \bar q_i}({\hat \tau \over {y_a y_b}},{q^2 \over \mu^2})
{}~f_{\Delta q_i /q}(y_b,\mu^2) +
\nonumber\\
& &{\sum_{i}} \int
_{\hat \tau}^{1} {dy_a \over y_a} \int_{\hat{\tau} \over y_a}^1 {dy_b \over
y_b}
{}~f_{\Delta q_i /q}(y_a,\mu^2)
{}~\widetilde H_{q_i g}({\hat \tau \over {y_a y_b}},{q^2 \over \mu^2})
{}~f_{\Delta g /g}(y_b,\mu^2),
\label{eq:E15}
\end{eqnarray}
where $m$ is the mass of the quark, $k^2$ is off-shell mass of the gluon
and $\hat \tau(\equiv \tau /x_a x_b)$ is the DY variable in the parton level.
The only unknown in eq.(\ref{eq:E15}) is $f_{\Delta \bar q/g}$ to order
$\alpha_s$, which is evaluated using eq.(\ref{eq:E7})
in the dimensional regularisation scheme.
The only contribution to
$f_{\Delta \bar q /g(h)}$ comes from the cut triangle diagram (fig. 2) with
quark and
antiquark loops.  The matrix element is only UV divergent if we keep
quarks on shell and gluon off-shell.  On regulating we get
\begin{eqnarray}
f_{{\Delta \bar q_i} \over {g(h)}}(\hat {\tau},\mu_R^2) &=& h {\alpha_s
\over {4\pi}} \left[(2\hat {\tau}-1)~ln{{\mu_R^2} \over {m^2-k^2\hat
{\tau}(1-\hat {\tau})}}-1+ {m^2\over {m^2-k^2 \hat {\tau}(1-\hat {\tau})}}
\right],
\label{eq:E16}
\end{eqnarray}
where $\mu _R^2=\mu^2 exp(\gamma_E+{2\over \epsilon})$ and $h$ is the helicity
of the gluon target.
Since $\widetilde H_{q \bar q}$ is given in eq.(\ref{eq:E14}), $\widetilde
H_{q g}$ can be found if $\widetilde W_{qg}$ is calculated.
As the compton
subprocess diagram (fig.3) has only mass singularities, we keep all the masses
non zero.  Neglecting the power corrections in the large $q^2$ and $s$
limit, we get
\begin{eqnarray}
\widetilde W_{qg}(\hat {\tau},q^2) & = & {\sum _i}
{\alpha_s}{e_i^2\over 6}  \left[ (2 \hat {\tau} -1)~ln{{q^2 (1- \hat {\tau})^2}
\over { \hat {\tau} (m^2-k^2 \hat {\tau} (1- \hat {\tau} ))}}-{1 \over 2}(3
\hat
{\tau} ^2 +2 \hat {\tau} -3) + \right.
\nonumber\\
& &  \left. {m^2 \over {m^2-k^2 \hat {\tau} (1- \hat {\tau} )}}  \right].
\label{eq:E17}
\end{eqnarray}
Now we can evaluate the hard scattering coefficient, by using the above
results
\begin{eqnarray}
\widetilde H_{qg}(\hat {\tau} ,q^2) & = & - \sum_{i} \alpha_s {e_i^2 \over 6}
\left[(2\hat {\tau} -1)~(ln{\hat {\tau} \over (1-\hat {\tau} )^2}+
ln{\mu^2_R \over {q^2}})+ {1 \over 2} (3 \hat {\tau} ^2 +
2 \hat {\tau} -3) -1 \right],
\nonumber\\
& = & \widetilde H_{\bar {q}g}(\hat {\tau} ,q^2).
\label{eq:E18}
\end{eqnarray}
We find that a similar analysis of the weak DY processes also yields the
same quark and gluon hard scattering coefficient functions but for weak
coupling factors.
Note that {\it the above expression is free of any mass singularities,
verifying the proposed factorisation to order $\alpha_s$ in the gluonic
sector}.  In addition, since the first moment of $\widetilde H_{qg}(\hat
\tau,q^2)$ vanishes, {\it there is no gluonic contribution to the first
moment of DY function through the `compton' subprocess for both
electromagnetic and weak processes}.

In the {\it parton model} analysis of the above problem, the calculation of
gluonic coefficient function in the DY function involves only
$\widetilde W_{qg}(\hat \tau ,q^2)$ term.  This being mass singular, a
particular choice of prescription has to be made to regulate these
singularities.  Depending on the prescription one gets different
results: for $m^2 \not= 0$, $k^2 = 0$ the first moment of $\widetilde
W_{qg}(\hat \tau , q^2)$ is zero, while for $m^2 = 0$, $k^2 \not= 0$ we
have,
\begin{eqnarray}
\int d \hat \tau \widetilde W_{qg} (\hat \tau ,q^2) & = &
-{1\over 2} \left ({\alpha_s e_i^2} \over 3 \right ).
\label{eq:E19}
\end{eqnarray}
It is interesting to note that {\it one arrives at the same conclusion even
in the polarised DIS case}.  Invoking universality of parton distributions
we expect the first moment of $\widetilde W_{P_1 P_2}(\tau, q^2)$ in
the parton model using the prescription $m^2 = 0$ , $k^2 \not= 0$, to be
of the form $\Delta q^{\prime}~\Delta \bar q^{\prime}$, where $\Delta
q^{\prime} = \Delta q~-~(\alpha_s /{4 \pi})\Delta g$ \cite{AR}.  Note
that eq.(\ref{eq:E19}) verifies this to order $\alpha_s$.  This implies
that the first moment of $\widetilde W_{gg}(\hat \tau, q^2)$ is
$(4 \pi e_i^2 /3)~(\alpha_s /4 \pi)^2$.  Assuming this to be true in
the proposed
factorisation scheme, one finds that the first moment of $\widetilde
H_{gg}(\hat \tau, q^2)$ is also zero.  Explicit computation of
$\widetilde H_{gg}$ involves evaluation of eight diagrams and three body
phase space integral.  We will be presenting these details elsewhere
and expect that it will corroborate that the first moment vanishes.
Hence, in the factorisation scheme, gluons can not contribute to the
first moment of polarised DY function.

We have studied the polarised DY process using in the factorisation method
and provide operator definitions for the polarised parton distribution
functions, which are consistent with that of DIS, hence confirms
universality.  We have explicitly shown the factorisation of mass
singularities to order $\alpha_s$ in the gluonic sector.  The gluonic
contribution to the first moment of the polarised DY function from the
`compton' subprocess turns out to be zero.  Invoking universality we
conclude that the gluonic contribution to the first moment of polarised
DY function vanishes.

We are grateful to G. Date, D. Indumathi, M.V.N. Murthy, R.
Ramachandran and S. Umasankar for useful discussions and suggestions.
Further we thank DI, MVNM, RR and SU for carefully reading the manuscript
and making useful suggestions.

\eject

\eject
{\Large \bf Figure Captions}
\begin{list}
{}{\setlength{\labelwidth}{20mm}}
\item [Fig 1.] Quark Born diagram.
\item [Fig 2.] Triangle diagrams contributing to polarised distribution
functions $f_{\Delta q/g(h)}$.
\item [Fig 3.] `Compton' subprocess.
\end{list}
\eject
\end{document}